\documentclass[epj,nopacs]{svjour}
\usepackage{psfig}

\begin{document}
\def\ackname{Acknowledgements.}
\def\sectcounterend{}
\def\arraystretch{1.2}
\def\ab{\allowbreak}
\def\hb{\hfill\break}

\overfullrule=0pt
\parskip=0pt

\title{\boldmath Another look at anomalous $J/\Psi$ suppression\\
in ${\mathrm {Pb + Pb}}$ collisions at $P/A = 158\,{\mathrm {GeV}}/c$}
\titlerunning{Another look at anomalous $J/\Psi$ suppression
in ${\mathrm {Pb + Pb}}$ collisions at $P/A = 158\,{\mathrm {GeV}}/c$}

\author{J. Gosset\thanks{e-mail: {Jean.Gosset@cea.fr}},
 A. Baldisseri,
 H. Borel,
 F. Staley,
 Y. Terrien}

\authorrunning{J. Gosset et al.}

\institute{DAPNIA/SPhN, CEA/Saclay, 91191 Gif-sur-Yvette Cedex,
France}

\date{Received: 7 June 1999 / Revised version: 13 September 1999 /\\
Published online: 3 February 2000 -- \copyright\ Springer-Verlag 2000}

\abstract{
A new data presentation is proposed in order to consider anomalous
$J/\Psi$ suppression in ${\mathrm {Pb + Pb}}$
collisions at $P/A=158$\,GeV/$c$.
If the inclusive differential cross section
with respect to a centrality variable is available,
one can plot the yield of $J/\Psi$ events per Pb--Pb collision
as a function of an estimated squared impact parameter.
Both quantities are raw experimental data and have a clear physical
meaning.
As compared to the usual $J/\Psi$ over Drell--Yan ratio,
there is a huge gain in statistical accuracy.
This presentation could be applied advantageously
to many processes in the field of nucleus--nucleus collisions
at various energies.
} 

\maketitle

\section{\boldmath Introduction}
\label{sec:intro}
Very interesting results have been obtained recently
by the NA50 experiment at CERN
concerning $J/\Psi$ production in ${\mathrm {Pb + Pb}}$ collisions
at  $P/A= 158$\,GeV/$c$ \cite{ABR97,ABR97a,RAM98,ROM98,ABR99}.
In most central collisions, the $J/\Psi$ events are significantly
suppressed
with respect to what is expected from normal nuclear absorption
as measured in lighter systems \cite{ABR97a,RAM98,ROM98,ABR99}.
According to theoretical predictions made more than ten years
ago
by Matsui and Satz \cite{MAT86},
this anomalous $J/\Psi$ suppression could be a sign of the awaited
formation
of a quark--gluon plasma in nucleus--nucleus collisions at very
high
energy.

The deficiencies of customary data presentations,
using the ratio between $J/\Psi$ and Drell--Yan events
or the differential cross section for $J/\Psi$ production
with respect to some measured centrality variable,
will be stressed first.
A new data presentation \cite{GOS99} will then be proposed,
which removes the previous deficiencies
with the help of one key additional ingredient,
the inclusive differential cross section
with respect to the centrality variable.
For any process ``p'', the yield of ``p'' events per nucleus--nucleus
collision
is plotted as a function of an estimated
squared impact parameter.
This new presentation will be applied to NA50 results
from their 1995 data taking, available in a thesis  \cite{BEL97}.
Implications of this new presentation will also be discussed.
Finally, conclusions will be drawn in the perspective
of RHIC and LHC experiments on nucleus--nucleus collisions at
very high energy.

\newpage

\section{\boldmath Usual data presentation}
\label{sec:usual}
In the usual presentation of NA50 results concerning $J/\Psi$
production
in nucleus--nucleus collisions \cite{ABR97,ABR97a,RAM98,ROM98},
the ratio between $J/\Psi$ and Drell--Yan events
is plotted as a function of the transverse energy $E_{\mathrm
{T}}$
measured in an electromagnetic calorimeter.
This centrality variable is aimed primarily at sorting out
all events according to  the impact parameter of the collision.
The Drell--Yan process, supposed to be insensitive to nuclear
matter effects,
is indeed a good reference
for normalizing $J/\Psi$ events from the physics point of view.
Moreover, some systematic effects cancel in such a ratio.
However, when one sees an interesting feature in a ratio,
it is not always obvious to know whether it is due to the numerator
or the denominator.
More importantly, since the Drell--Yan continuum
is much less populated than the $J/\Psi$ peak in the dimuon mass
spectrum,
the statistical uncertainty on the ratio
comes essentially from the denominator.
It is between 5 and 10 times larger
-- 5 for central collisions, 10 for peripheral ones --
than the contribution from the number of events in the $J/\Psi$
peak.
In other words, one would need between 25 and 100 times less
running time,
all other conditions staying equal,
to get a given relative statistical uncertainty on $J/\Psi$ production
from the number of events in the $J/\Psi$ peak
than from the ratio between $J/\Psi$ and Drell--Yan events.
This is a first deficiency of the usual presentation.
One would like to use another quantity for normalizing $J/\Psi$
events
without losing so much in statistical accuracy.

The ratio between $J/\Psi$ and Drell--Yan events
is obtained from the differential $E_{\mathrm {T}}$ distributions
of cross section for both classes of events,
which are the basic experimental data one has to start with.
These raw experimental data, $d \sigma_{\mathrm {p}} / d E_{\mathrm
{T}}$ and
$E_{\mathrm {T}}$,
where ``p'' stands for either $J/\Psi$ or Drell--Yan process,
do not have a very direct physical meaning.
From the increase or the decrease of
$\mathrm {d} \sigma_{\mathrm {p}}/\mathrm {d}_{\mathrm {T}} $
as a function of $E_{\mathrm {T}}$, one cannot even infer
whether the production of ``p'' events
increases or decreases with the centrality of the collision.
In particular, the decrease of
$\mathrm {d} \sigma_{\mathrm {p}} / \mathrm {d} E_{\mathrm {T}}$
at high $E_{\mathrm {T}}$ for any ``p'' process
simply reflects the fact that, for most central collisions,
there is a maximum value of $E_{\mathrm {T}}$
beyond which there is no more cross section.
$E_{\mathrm {T}}$ is surely increasing with centrality,
but at which rate?
The answer to this question is needed if one wants to go from
$E_{\mathrm {T}}$
to a more direct centrality variable like the impact parameter.
For all these reasons, ${\mathrm {d}} \sigma_{\mathrm {p}} /
{\mathrm {d}} E_{\mathrm
{T}}$ and $E_{\mathrm {T}}$
are rather difficult to understand
and compare directly with simple models.
This is the second deficiency of the usual presentation.
One would like to find other experimental quantities,
not too far from $\mathrm {d} \sigma_{\mathrm {p}} / \mathrm
{d} E_{\mathrm {T}}$
and $E_{\mathrm {T}}$,
which would have a more direct physical meaning,
from which one could directly say something
on the variation of the production of ``p'' events with centrality,
and which could be compared directly with simple models.

\section{\boldmath New data presentation}
\label{sec:new}
One key quantity that could be used to remove the above-mentioned
deficiencies
is the inclusive distribution of the cross section for
the centrality variable, denoted as $C$ hereafter for more generality.
This inclusive distribution will be needed in its differential
form
$\mathrm {d} \sigma_{\mathrm {inc}} / \mathrm {d} C$ for normalizing
$\mathrm {d} \sigma_{\mathrm {p}} / \mathrm {d} C$,
and in its integral form $\sigma_{\mathrm {inc}}(C)$
for getting an estimated squared impact parameter $(b^2)_{\mathrm
{e}}
$.

When we use $\mathrm {d} \sigma_{\mathrm {p}} / \mathrm {d} C$
we mix the probability to get a given centrality
with the probability to get the ``p'' process at this centrality.
This is precisely why
the increase or the decrease of $\mathrm {d} \sigma_{\mathrm
{p}}
/ \mathrm {d} C$
as a function of centrality has no straightforward meaning.
This distribution
$\mathrm {d} \sigma_{\mathrm {p}} / \mathrm {d} C$ is in fact
the product of two quantities
which themselves have a more direct physical meaning than their
product.
It can be written as
$Y_{\mathrm {p}}\cdot {\mathrm {d}} \sigma_{\mathrm {inc}} /
\mathrm {d} C$,
where the inclusive distribution ${\mathrm {d}} \sigma_{\mathrm
{inc}} /
{\mathrm {d}} C$
carries the probability that a nucleus--nucleus collision occurs
at a given value $C$ of the centrality variable,
and $Y_{\mathrm {p}}$ is the yield of ``p'' events per nucleus--nucleus
collision
at this given centrality.
This yield $Y_{\mathrm {p}}$, which, being
equal to ($\mathrm {d} \sigma_{\mathrm {p}} / \mathrm {d}
C)/\-({\mathrm {d}} \sigma_{\mathrm
{inc}} / {\mathrm {d}} C$),
is a well-defined physical quantity.
For copiously produced particles
it is simply their average multiplicity per nucleus--nucleus
collision
at a given centrality.
As it is a ratio, it should be insensitive to some systematic
uncertainties.
Its variation as a function of $C$
should accurately reflect
whether the production of ``p'' events
increases or decreases with the centrality of the collision.
It should not be subject to any artificial decrease
for most central collisions.
For both $J/\Psi$ and Drell--Yan processes,
one expects that this yield steadily increases towards more central
collisions,
like the number of nucleon--nucleon collisions they originate
from,
unless the $J/\Psi$ is very strongly suppressed.
The first step in the new data presentation
is thus to use the yield $Y_{\mathrm {p}}$ of ``p'' events
per nucleus--nucleus collision instead of ${\mathrm {d}} \sigma_{\mathrm
{p}} / {\mathrm {d}} C$.

The idea behind tagging a process with a centrality variable
in nucleus--nucleus collisions is always to sort out events
according to the impact parameter.
If a centrality variable $C$ is assumed to vary monotonically
as a function of the impact parameter $b$
-- and centrality variables are purposely chosen for that reason
--,
it is very easy to go from $C$ to an estimate of $b$,
or more precisely $b^2$.
One only has to use the integral inclusive cross section $\sigma_{\mathrm
{inc}}
(C)$,
from most central collisions to any given value of $C$.
From the geometrical dependence of the inclusive cross section,
${\mathrm {d}}\sigma_{\mathrm {inc}}=2{\pi}{\cdot}b\cdot
{\mathrm {d}} b=\pi\cdot {\mathrm {d}} (b^2)$,
one simply gets
$\sigma_{\mathrm {inc}}(C)=\pi(b^2)_{\mathrm {e}}$
where $(b^2)_{\mathrm {e}}$ is an estimate of the squared impact
parameter
corresponding to the value of $C$.
There is a one-to-one correspondence between $C$ and $(b^2)_{\mathrm
{e}}
$.
$C$ slices are transformed into $(b^2)_{\mathrm {e}}$ slices
with a width proportional to the number of counts in the $C$
slices.
For this reason, which is also related to the fact that
$\mathrm {d} \sigma_{\mathrm {inc}} / \mathrm {d} b$ goes to
zero at zero impact
parameter,
$(b^2)_{\mathrm {e}}$ seems a better variable than
the estimated impact parameter $b_{\mathrm {e}}$.
Instead of dividing $\sigma_{\mathrm {inc}}(C)$ by $\pi$,
one could divide it by the geometrical cross section $\sigma_{\mathrm
{geo}}
$
and get a quantity proportional to $(b^2)_{\mathrm {e}}$,
but with such a normalization that it varies between 0 and 1
from most central to most peripheral collisions.
Another quantity which could be interesting to use
for plotting results from different systems
would be $(b_{\mathrm {max}}^2-(b^2)_{\mathrm {e}})$,
where $b_{\mathrm {max}}^2=\sigma_{\mathrm {geo}}/\pi$.
It has the advantage of being correlated, and not anticorrelated,
with the centrality, and extends to larger and larger values
for larger and larger systems,
with the zero value corresponding always to most peripheral collisions.
Such a transformation from $\sigma_{\mathrm {inc}}(C)$ to $(b^2)_{\mathrm
{e}}
$
has been used in more or less details
by several experiments
in the field of nucleus--nucleus collisions
at various incident energies.
Its reliability has been discussed thoroughly,
and has been checked to be excellent within the framework
of the intranuclear cascade model
at energies per nucleon around 1\,GeV \cite{CAV90}.
Model calculations are obviously needed to evaluate the method
and to compare the quality factors
of various centrality variables \cite{CUG83},
which combine
the fluctuations of $C$ at any given $b$
and the variation rate of $C$ with respect to $b$ or $b^2$.
For some AGS or SPS experiments,
even though the data are plotted as a function of a centrality
variable,
a scale for the impact parameter estimated along the preceding
lines
is indicated in parallel \cite{AGG98,BAR99}.
Finally, the second step in the new data presentation
consists in replacing the measured centrality variable $C$
with an estimate of the squared impact parameter,
$(b^2)_{\mathrm {e}} = \sigma_{\mathrm {inc}}(C)/\pi$.

After applying both steps one gets
the yield $Y_{\mathrm {p}}$ of ``p'' events per nucleus--nucleus
collision
as a function of the estimated squared impact parameter.
Both quantities are raw experimental data and have a clear physical
meaning.
As compared with the usual ratio between $J/\Psi$ and Drell--Yan
events,
there is a huge gain in statistical accuracy for $J/\Psi$.

With this new data presentation
one can consider any process independently of all others,
with the best statistics available for each of them.
In the same way as
the integral of
$\mathrm {d} \sigma_{\mathrm {p}} / \mathrm {d} C$ as a function
of $C$
is the total cross section for the ``p'' process,
the integral of $Y_{\mathrm {p}}$ as a function of $(b^2)_{\mathrm
{e}}$
is equal to the total cross section for the ``p'' process divided
by $\pi$.
There is no loss of information in going from the centrality
variable $C$
to the estimated squared impact parameter $(b^2)_{\mathrm {e}}$.
The limits of the slices used for looking at the variation with
centrality
have simply to be specified for both $C$ and $(b^2)_{\mathrm
{e}}$.
The main requirement for this new presentation
is a good inclusive centrality distribution,
with high enough statistical accuracy
and proper corrections for efficiency and empty-target contribution.
The yield $Y_{\mathrm {p}}$ being the ratio of cross sections,
part of systematic uncertainties are removed
if inclusive measurements are taken simultaneously with
the measurements of the ``p'' process.
The normalization uncertainty of the inclusive cross section
has an effect on the abscissa rather than on the ordinate,
which may be unusual but does not bring about any practical problem.

Such a presentation provides an excellent starting point
for comparisons between experiment and theory,
and also between experimental results themselves.
Since there is no explicit appearance of $C$ in the new presentation,
results obtained with various centrality variables
should be identical
as long as these variables
sample the impact parameter the same way.
Anyway, such a comparison could help to check systematic uncertainties.
For comparisons between experiment and theory,
it is straightforward to compare
the measured yields as a function of the estimated $b^2$
with the calculated ones as a function of the real $b^2$.
This is particularly interesting for a quick comparison with
simple
models.
However, in order to take into account the fluctuations
of any centrality variable as a function of the impact parameter,
a comparison with better quality would result from using the
same procedure
of $b^2$ estimation for both experiment and theory,
even if inclusive cross sections do not agree
within a high degree of accuracy \cite{CAV90},
or from unfolding the experimental results from these fluctuations.
It is also clear that this whole presentation could be applied
advantageously to other processes than $J/\Psi$ and Drell--Yan
production.

\section{\boldmath Application}
\label{sec:appl}
\begin{figure}
\psfig{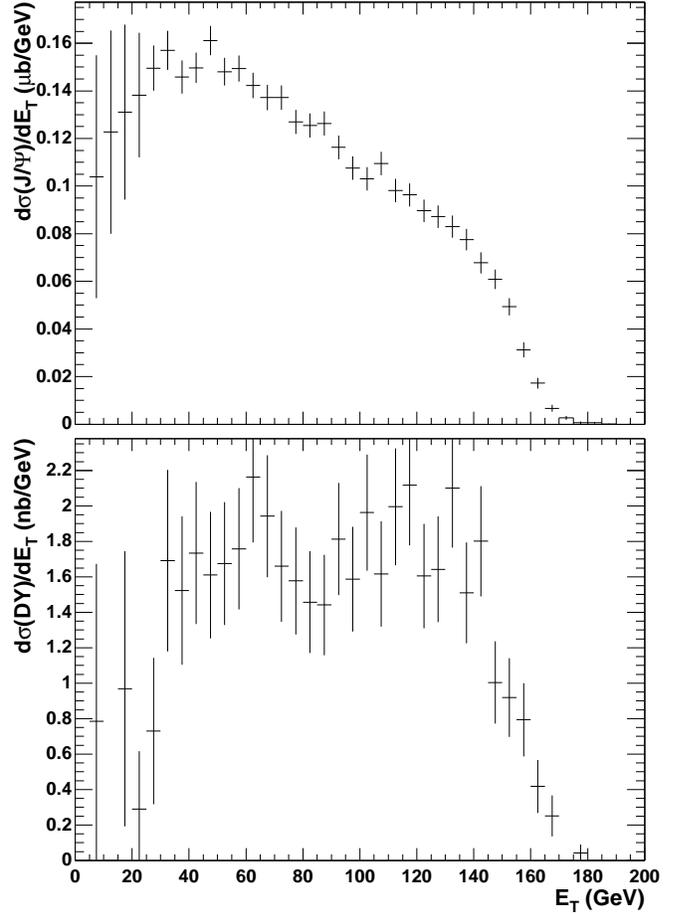}
\caption[]{Differential transverse-energy distribution of cross
section
for $J/\Psi$ (top) and Drell--Yan (bottom) production
in Pb--Pb collisions at $P/A=158$\,GeV/$c$ \protect\cite{BEL97}
\label{fig:basic_data_PsiDY}}
\end{figure}

\begin{figure}
\psfig{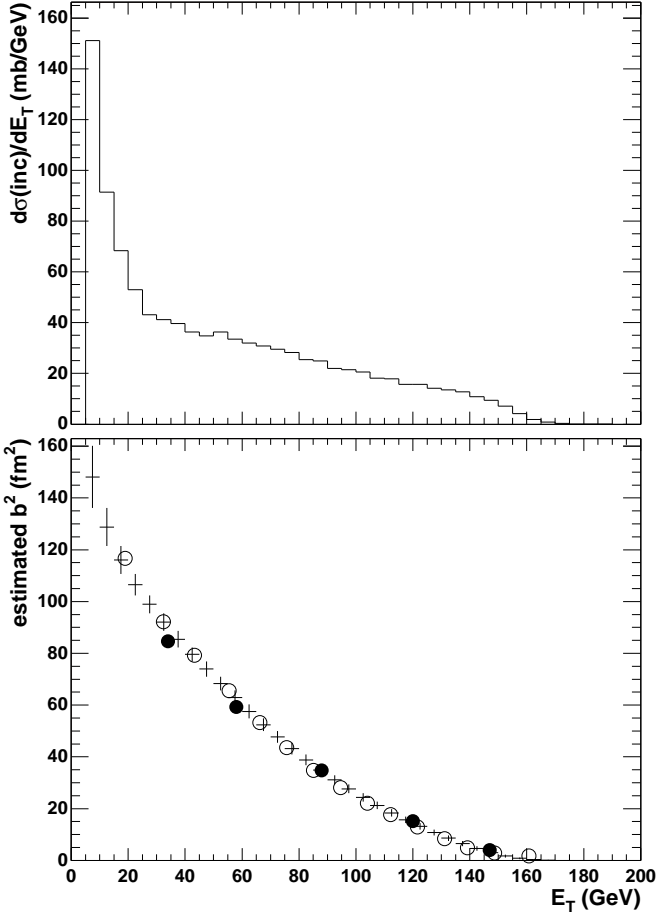}
\caption[]{Top: differential transverse-energy distribution
of inclusive cross section
in Pb--Pb collisions at $P/A=158$\,GeV/$c$ \protect\cite{BEL97}.
Bottom: estimated squared impact parameter
versus transverse energy;
the correspondence between $E_{\mathrm {T}}$ and $b$
from \protect\cite{ABR97,ABR99} is
indicated with closed and open circles, respectively
\label{fig:basic_data_inc}}
\end{figure}

In a thesis by Bellaiche \cite{BEL97} from the NA50 collaboration,
all necessary pieces of information are available from the 1995
data taking
for applying this new data presentation.
They have not been published as such.
The results presented below thus have to be considered with care.
They are only indicative,
and they are to be used simply as an illustration of the advantages
inherent to the new data presentation.
Basic experimental data are the
differential $E_{\mathrm {T}}$ distributions of the cross section
for the $J/\Psi$ and Drell--Yan events (Fig.~\ref{fig:basic_data_PsiDY}).
The key additional ingredient is
the differential $E_{\mathrm {T}}$ distribution of the inclusive
cross
section,
by the integration of which one can estimate
the squared impact parameter $(b^2)_{\mathrm {e}}$
for each value of $E_{\mathrm {T}}$ (Fig.~\ref{fig:basic_data_inc}).
Uncertainties on $\mathrm {d} \sigma_{\mathrm {inc}} / \mathrm
{d} E_{\mathrm {T}}$
have been neglected in the following.
The inclusive cross section is only available
with arbitrary units in the thesis.
A normalization factor had to be introduced to get the $(b^2)_{\mathrm
{e}}
$ values
in units of fm$^2$.
It has been adjusted in such a way that the dependence of $(b^2)_{\mathrm
{e}}
$
upon $E_{\mathrm {T}}$ agrees with the correlation between
the average values of $E_{\mathrm {T}}$ and $b$
listed in NA50 publications \cite{ABR97,ABR99}
for successive $E_{\mathrm {T}}$ slices,
as fitted on the basis of a Glauber model calculation.
These average values are also shown in the bottom part of
Fig.~\ref{fig:basic_data_inc}.
$E_{\mathrm {T}}$ values from \cite{ABR99}
have been divided by 0.74 to take into account
the different $E_{\mathrm {T}}$ scales used in the NA50 publications.

\begin{figure}
\psfig{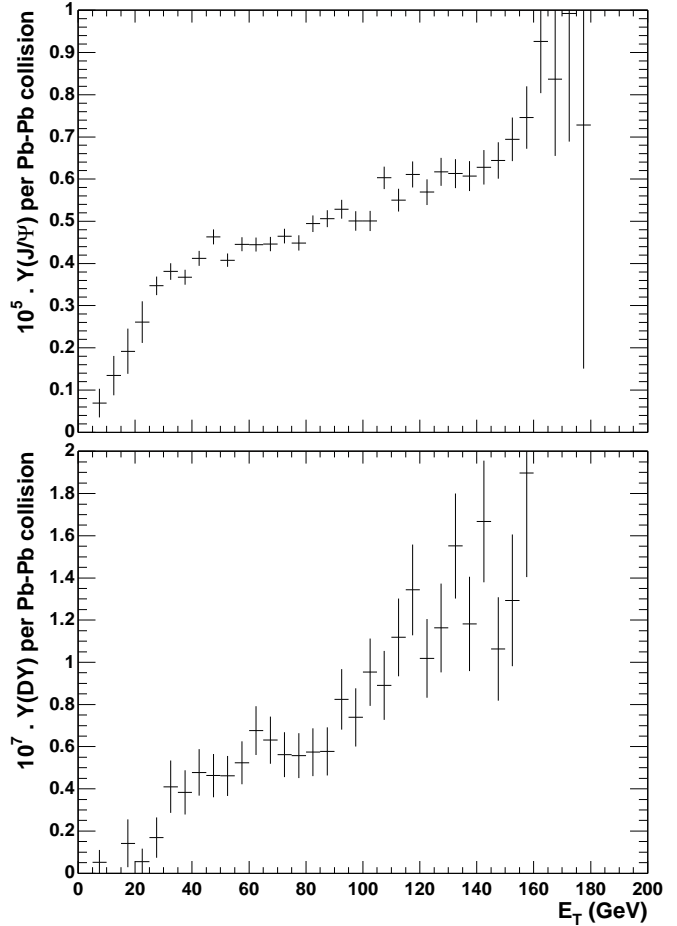}
\caption[]{Yields of $J/\Psi$ (top) and Drell--Yan (bottom) events
per Pb--Pb collision
versus transverse energy,
calculated from NA50 data at $P/A=158$\,GeV/$c$ \protect\cite{BEL97}
\label{fig:first_step}}
\end{figure}

After the first step,
i.e. the normalization of the $J/\Psi$ and Drell--Yan cross sections
to the inclusive one,
one gets (Fig.~\ref{fig:first_step})
the yields of $J/\Psi$ and Drell--Yan events
per Pb--Pb collision as a function of $E_{\mathrm {T}}$.
Both yields increase with $E_{\mathrm {T}}$,
without any artificial decrease at large $E_{\mathrm {T}}$.
Whereas this increase is rather steady for Drell--Yan events,
there is a change of behaviour for $J/\Psi$ at about 40\,GeV,
an $E_{\mathrm {T}}$ value
beyond which the increase is definitely slower
for $J/\Psi$ than for Drell--Yan events.
This is an indication for $J/\Psi$ suppression
in central collisions, relative to Drell--Yan events.

\begin{figure}
\psfig{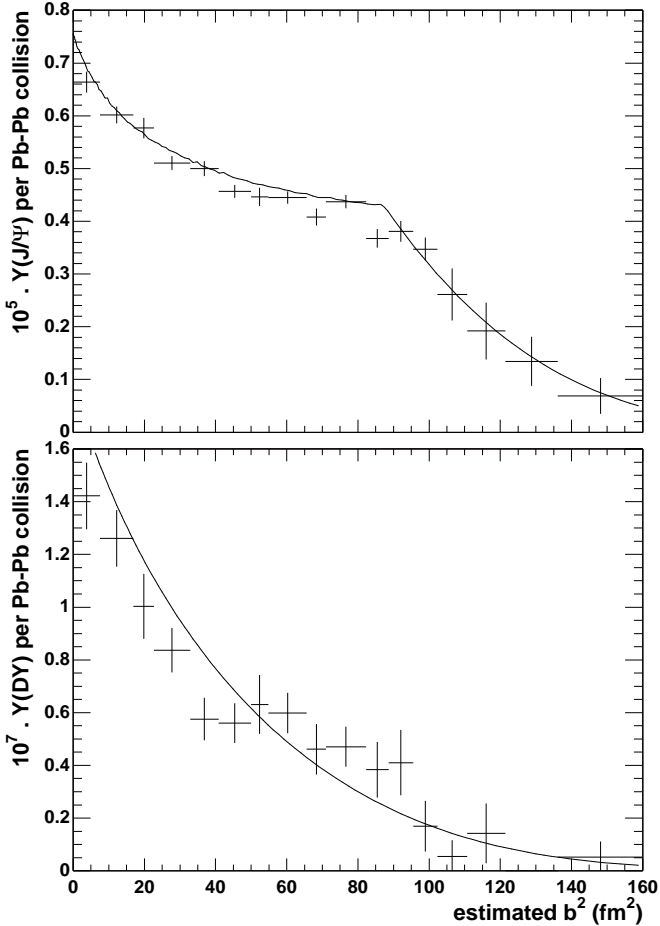}
\caption[]{Yields of $J/\Psi$ (top) and Drell--Yan (bottom) events
per Pb--Pb collision
versus estimated squared impact parameter,
calculated from NA50 data at $P/A=158$\,GeV/$c$ \protect\cite{BEL97}
(crosses) and from the number of nucleon--nucleon collisions
within a Glauber model (solid lines), with correction factors
for $J/\Psi$ absorption
\protect\cite{BLA96} (see text for more details)
\label{fig:second_step}}
\end{figure}

After the second step,
i.e. replacing $E_{\mathrm {T}}$ by $\sigma_{\mathrm {inc}}(E_{\mathrm
{T}})/\pi$,
one gets (Fig.~\ref{fig:second_step})
the yields of $J/\Psi$ and Drell--Yan events
per Pb--Pb collision as a function of $(b^2)_{\mathrm {e}}$,
the squared impact parameter estimated from the
$E_{\mathrm {T}}$ inclusive cross section.
Slices with almost constant width in $(b^2)_{\mathrm {e}}$ have
been
used.
The same remarks could be made as from Fig.~\ref{fig:first_step}
after the first step.
The $E_{\mathrm {T}}$ value of 40\,GeV for the change of behaviour
of the $J/\Psi$ yield is changed into a $(b^2)_{\mathrm {e}}$
value
of 80\,fm$^2$.
Points with large error bars at large $E_{\mathrm {T}}$
in Fig.~\ref{fig:first_step}
are all contained in the point at the smallest value of $(b^2)_{\mathrm
{e}}
$
in Fig.~\ref{fig:second_step}.
The limits of the yields for most central collisions
are more easily readable from Fig.~\ref{fig:second_step}
than from Fig.~\ref{fig:first_step}.
They could be directly compared with the $J/\Psi$ and Drell--Yan
yields
in p--p collisions
-- $J/\Psi$ and Drell--Yan cross sections divided by
the total inelastic p--p cross section --
times the number of nucleon--nucleon collisions
in most central Pb--Pb collisions
from a Glauber model calculation.
From integration of the yields in Fig.~\ref{fig:second_step}
one can get the total cross sections for $J/\Psi$ and Drell--Yan
production
divided by $\pi$.

A comparison is also made in Fig.~\ref{fig:second_step}
with a model calculation
\`a la Blaizot and Ollitrault \cite{BLA96}.
The yield of Drell--Yan events per Pb--Pb collision
as a function of $(b^2)_{\mathrm {e}}$
is compared, within a scale factor,
to the number of nucleon--nucleon collisions
calculated in the Glauber model
as a function of the real $b^2$,
without taking into account the fluctuations
between the estimated and actual $b^2$.
The agreement is reasonable.
For $J/\Psi$ events this number of nucleon--nucleon collisions
is multiplied by two correction factors for absorption.
The first one corresponds to normal $J/\Psi$ absorption
in nuclear matter with some cross section $\sigma_{\mathrm {abs}}$.
The second one is intended for simulating complete $J/\Psi$ suppression
due to quark--gluon plasma formation.
It goes down from one to zero
when nucleon--nucleon collisions producing $J/\Psi$ occur
in a tube of nuclear matter with nucleon density per unit area
larger than a critical value $\rho_{\mathrm {crit}}$.
With $\sigma_{\mathrm {abs}}=6.0$\,mb
and $\rho_{\mathrm {crit}}=2.9$\,fm$^{-2}$,
the model reasonably accounts for the experiment,
in particular for the clear change of behaviour at an
impact parameter of about 9\,fm.

Finally, very accurate results are obtained for $J/\Psi$ production
which can be compared easily with simple models.
The onset of the anomalous $J/\Psi$ suppression can be looked
at
with much better accuracy than
on the basis of the usual $J/\Psi$ over Drell--Yan
ratio.
However, we want to recall the word of caution
from the beginning of this section.
Definite conclusions about $J/\Psi$ anomalous suppression
need to be drawn from official data.
This is also why there was no attempt to calculate error bars
for the values of $\sigma_{\mathrm {abs}}$ and $\rho_{\mathrm
{crit}}$.
Moreover the Drell--Yan production remains an essential result.
One has to check its normal behaviour within its inherently limited
accuracy.

\section{\boldmath Discussion}
\label{sec:disc}
One idea from this new data presentation,
the normalization of $J/\Psi$ events
to the inclusive, or minimum bias, $E_{\mathrm {T}}$ distribution,
has been used recently by the NA50 collaboration \cite{ABR99},
making the most of the whole statistics available for $J/\Psi$
production
in their 1996 data taking.
However, this new presentation is not used as such,
except the first step for Drell--Yan production only.
For easy comparison with previously published results,
the $J/\Psi$ yield is transformed into a
``minimum bias'' $J/\Psi$ over Drell--Yan ratio,
through a division by a model calculation for the Drell--Yan
yield.
In order to stick more closely to the raw experimental data,
it would be very interesting if the new presentation
were to be applied as a whole to these most recent and also to
future
NA50 data.
One would not have to worry anymore
because of the different $E_{\mathrm {T}}$ scales
used in successive presentations.
More importantly,
it would be particularly helpful to compare between one another
the results obtained with the three centrality variables
available in this experiment, namely the transverse energy $E_{\mathrm
{T}}
$
measured in an electromagnetic calorimeter,
the zero-degree energy measured in a hadronic calorimeter,
and the multiplicity measured in a silicon detector.
Perhaps it would also be possible to get more accurate
information on $J/\Psi$ production from older data takings,
for example in ${\mathrm {S+U}}$ collisions.

\begin{figure}
\psfig{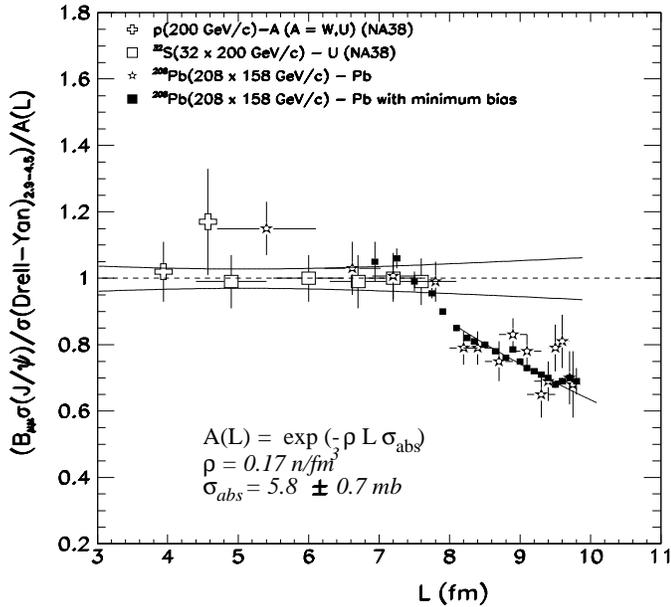}
\caption[]{``Minimum bias'' $J/\Psi$ over Drell--Yan ratio divided
by the normal
absorption suppression
as a function of the mean nuclear path length $L$
(Fig.~9 of \protect\cite{ABR99}.)
An uncertainty band for normal absorption,
and an exponential fit for additional absorption in most central
collisions
beyond $L=8$\,fm
have been added to the original figure
\label{fig:absorption_band}}
\end{figure}

Finally, it is interesting to try and quantify the gain
brought about by the normalization to the inclusive centrality
distribution
in the assessment of the anomalous $J/\Psi$ suppression in ${\mathrm
{Pb + Pb}}$
collisions.
It can be done for example on NA50 results as shown
in Fig.~9 from \cite{ABR99}.
In this figure, the ``minimum bias'' as well as the measured
$J/\Psi$ over Drell--Yan ratios
are divided by the normal absorption factor and plotted
as a function of the mean nuclear path length $L$
(Fig.~\ref{fig:absorption_band}).
The normal absorption appears as a horizontal line
at a constant value of 1,
without any information on its uncertainty.
In Fig.~\ref{fig:absorption_band}, an uncertainty band,
necessary for a quantitative comparison to ${\mathrm {Pb + Pb}}$
results, has
been added
around the straight reference line.
It has been calculated from the same p nucleus and ${\mathrm {S+U}}$
data
as used in \cite{ABR99},
with correct error bars as compared to previous NA50 publications
(see note added in proof to \cite{RAM98}),
and taking into account the correlation between
the normalization and the slope of the exponential fit.
An uncertainty band had already been shown in \cite{KHA97}
but it had been calculated with the old error bars for ${\mathrm
{S+U}}$
data
and without taking into account the correlation between
the normalization and the slope.
By chance this uncertainty band was not too much wrong
since both effects were roughly compensating for each other.
One way to quantify the discrepancy of ${\mathrm {Pb + Pb}}$
data
from normal nuclear absorption
consists in fitting the points corresponding to most central
collisions,
i.e. beyond $L=8$\,fm,
with an exponential function of $L$ (Fig.~\ref{fig:absorption_band}).
The result is an effective additional absorption cross section
of 9\,mb,
with uncertainties of 0.6 and 2.6\,mb
depending on whether one uses the ``minimum bias''
or the measured $J/\Psi$ over Drell--Yan ratio.
With respect to the reference absorption cross section of $ 5.8\pm
0.7$\,mb,
the significance of this additional absorption
amounts to 9.8 or 3.3 standard deviations, respectively,
with a clear advantage to the ``minimum bias'' ratio,
because it uses the normalization
to the inclusive centrality distribution.

\section{\boldmath Conclusion and perspectives}
\label{sec:conc}
A new data presentation has been proposed
for results from nucleus--nucleus collisions.
Its domain of application is not limited to
$J/\Psi$ and Drell--Yan production processes
which have been chosen for illustration.
For any ``p'' process one ends
with its yield per nucleus--nucleus collision
as a function of the estimated squared impact parameter.
Since the normalization of the yield refers to the most probable,
i.e. inclusive, process,
one keeps the best statistical accuracy for each process.
It seems to be a good way for going as far as possible
with raw experimental data,
sticking as closely as possible to them
while trying to show physical quantities of interest.
Could it be the best way to present experimental
results concerning nucleus--nucleus collisions
before comparison to any model?
From the theoretical side, one would like to compare experimental
results
with results from model calculations on plots using
the most relevant variable from the model,
for instance the number of participants,
the number of nucleon--nucleon collisions,
the mean path length in nuclear matter, etc.
The proposed data presentation
could serve as the common basis
before going to any of these plots.

For future nucleus--nucleus experiments that will take place
at
RHIC and LHC,
the present work shows that it is possible to study any process
independently of all others.
The experimental results to be presented for each process
have a direct physical meaning
and are easily compared with model calculations.
The only requirement is the measurement
of the inclusive differential cross section
with respect to at least one centrality variable
used to sort out events according to impact parameter.
It is essential that such inclusive measurements be available
in experiments to be performed at RHIC and LHC.

\newpage

\begin{acknowledgement}
The authors want to express their thanks to F. Bellaiche
for providing them with the values of $J/\Psi$ and Drell--Yan
cross sections
from his thesis.
Discussions with J.-P. Blaizot, J. H\"ufner, J.-Y. Ollitrault
and H. Satz
are gratefully acknowledged.
\end{acknowledgement}

\end{document}